\documentclass[a4paper,aps,prl,notitlepage,superscriptaddress,longbibliography,twocolumn,10pt]{revtex4-2}

\usepackage{amsmath, amssymb, amscd, amsthm, amsfonts, physics}
\usepackage[utf8]{inputenc}
\usepackage{hyperref}
\usepackage{subcaption}
\usepackage[dvipsnames]{xcolor}
\usepackage[mathscr]{euscript}
\usepackage{graphics,graphicx,float,xfp}
\usepackage{braket}
\usepackage{epstopdf}

\usepackage{pgfplots}
\pgfplotsset{compat=1.17}
\usepackage[normalem]{ulem}
\usepackage[compat=1.1.0]{tikz-feynman}

\usetikzlibrary{calc,patterns,angles,quotes,math,decorations.text,arrows.meta,decorations.markings}
\tikzstyle{world line}=[blue!70!black,line width=0.4]
\tikzstyle{world line thick}=[blue!80!black,line width=3]
\tikzstyle{world line t}=[purple!60,line width=0.4]

\tikzset{declare function={kruskal(\x,\c)  = {\fpeval{asin( \c*sin(2*\x) )*2/pi}};
}}
\tikzfeynmanset{warn luatex=false}

\begin{document}
\title{Hawking evaporation and the fate of black holes in loop quantum gravity}

\author{Idrus Husin Belfaqih}
\email{i.h.belfaqih@sms.ed.ac.uk}
\affiliation{Higgs Centre for Theoretical Physics, School of Physics and Astronomy, University of Edinburgh, Edinburgh EH9 3FD, UK}

\author{Martin Bojowald}
\email{bojowald@psu.edu}
\affiliation{Institute for Gravitation and the Cosmos, The Pennsylvania State
University,104 Davey Lab, University Park, PA 16802, USA}

\author{Suddhasattwa Brahma}
\email{suddhasattwa.brahma@gmail.com}
\affiliation{Higgs Centre for Theoretical Physics, School of Physics and Astronomy, University of Edinburgh, Edinburgh EH9 3FD, UK}

\author{Erick I.\ Duque}
\email{eqd5272@psu.edu}
\affiliation{Institute for Gravitation and the Cosmos, The Pennsylvania State
University,104 Davey Lab, University Park, PA 16802, USA}

\begin{abstract}
  A recent covariant formulation, that includes non-perturbative effects from
  loop quantum gravity (LQG) as self-consistent effective models, has revealed
  the possibility of non-singular black hole solutions. The new framework
  makes it possible to couple scalar matter to such LQG black holes and derive
  Hawking radiation in the presence of quantum space-time effects while
  respecting general covariance. Standard methods to derive particle
  production both within the geometric optics approximation and the
  Parikh--Wilczek tunneling approach are therefore available and confirm the
  thermal nature of Hawking radiation. The covariant description of scale-dependent decreasing holonomy corrections maintains
  Hawking temperature as well as universality of the low-energy transmission
  coefficients, stating that the absorption rates are proportional to the
  horizon area at leading order.  Quantum-geometry effects enter the thermal
  distribution only through sub-leading corrections in the greybody
  factors. Nevertheless, they do impact energy emission of the black hole and
  its final state in a crucial way regarding one of the main questions of
  black-hole evaporation: whether a black-to-white-hole transition, or a
  stable remnant, is preferred. For the first time, a first-principles derivation, based on a discussion of  backreaction, finds evidence
  that points to the former outcome.
\end{abstract}

\maketitle

\noindent {\it \textbf{Introduction}:} Hawking's celebrated black-hole
information loss paradox has been instrumental in fueling research in
fundamental physics over the last half-century. A black hole, which forms when
a sufficiently  massive star collapses onto itself, has an event horizon that
effectively isolates it from the rest of the universe. However, it was soon
understood that, quantum-mechanically, black holes behave as thermal objects,
emitting black-body radiation as viewed by an external observer.  
This observation immediately gives rise to a paradox. What happened to the
pure state of the star whose collapse formed the black hole if, in the end, it
completely evaporates by Hawking radiation? Two effects, ignored in Hawking's
calculation, can help resolve this question. First, backreaction of the
radiation stress-energy tensor on the spacetime geometry, although small for
large-mass black-holes, could become non-negligible at later stages of the
evaporation process as the mass shrinks to zero. Secondly, at the final stage,
backreaction alone is insufficient and quantum-gravity corrections must
be considered. Since Hawking evaporation in classical spacetime ends at a
singularity, complicating any discussion of backreaction effects on the
shrinking black hole, potential regularization effects from quantum gravity,
provided they are sufficiently well-understood and preserve a meaningful and
covariant spacetime interpretation, may make it possible to address these
questions. This is our primary motivation.

Vacuum spherically-symmetric solutions of loop quantum gravity (LQG) have
hinted at resolving the singularity at the core of black holes by a quantum
`transition surface' \cite{alonso2022nonsingular, BBVeffBH,
  Idrus1}. Nevertheless, self-consistent dynamical solutions must be examined
to understand the fate of stellar collapse, and thus how the information-loss
paradox is resolved in LQG. In this Letter, we study Hawking radiation from a
covariant LQG black-hole model with scale-dependent decreasing holonomy
functions, a refined version of what is commonly referred to as the
\(\bar{\mu}\)-scheme \cite{APSII}. While we recover a thermal distribution with the same
asymptotic temperature as in general relativity, \(T_{\rm H} \sim 1/M\), LQG
parameters related to space-time discreteness affect the greybody factors, and
hence, the emission rate. This new effect provides a mechanism for thermal
stabilization and suggests, when taken in isolation, the formation of a
remnant at a sub-Planckian mass. However, horizon fluctuations as well as new
features of the quantum gravitational force at high density, that destabilize
the system gravitationally, must also be taken into account.  These crucial
ingredients are derived from a covariant spacetime description of LQG
modifications, combined with a qualitative discussion of backreaction in the
presense of matter couplings, as well as implications of the uncertainty
relation. The derived mass hierarchy of the three central effects --- thermal
stabilization, significant horizon fluctuations, and gravitational
destabilization --- shows that a black-to-white-hole transition is preferred
in LQG over the formation of a remnant. Our conclusion requires a careful
consideration of consistency conditions that have only recently become
available for spacetime models including quantum-geometry corrections. To
simplify the discussion, we use Planck units where $c=\hbar=k_{\rm B}=G=1$.

\noindent {\it \textbf{Effective LQG}:} Emergent modified gravity \cite{EMG,
  EMGCov} is a Hamiltonian approach to gravity that incorporates spacetime not
as a presupposed quantity but rather as an object derived (or emerging) from a
set of covariance conditions. This key idea results in weaker assumptions on
the basic setup of a gravitational system and evades previous no-go theorems
about possible covariant realizations of central LQG effects. As a result, new
theories are now available which, in particular, include modification
functions in spherically symmetric models that play the role of holonomy
corrections of LQG \cite{EMGCov,Idrus1}. In the standard Schwarzschild gauge,
solving the effective LQG equations of motion results in the line-element
\begin{equation}\label{static Schwarzschild}
    {\rm d}s^2
    = - f(x) {\rm d}t^2 +\frac{{\rm d}x^{2}}{h(x)f(x)}
    + x^2{\rm d}\Omega^2
\end{equation}
where $f(x)=1-2M/x$, $h(x)=1+\lambda(x)^{2}f(x)$, and $\lambda(x)$ is the
holonomy function with classical limit $\lambda \rightarrow 0$. Here, we
restrict discussions to a decreasing function $\lambda(x)=\sqrt{\Delta}/x$
where $\Delta \propto \ell_P^2$ may be interpreted as a discreteness parameter in LQG,
mimicking the traditional $\bar{\mu}$-scheme \cite{APSII}. (The analysis for a
general $\lambda(x)$ can be found in \cite{Idrus3}.)  The line-element
(\ref{static Schwarzschild}) preserves a horizon at the classical location
$x=2M=:x_{\rm H}$. It contains an additional coordinate singularity at
$x=x_{\Delta}<x_{\rm H}$ defined as the solution to $h(x_{\Delta})=0$ and given
by
\begin{eqnarray}
   x_{\Delta}=3M\delta^{1/3} \frac{\left( 1 + \sqrt{1+\delta}\right)^{2/3} - \delta^{1/3}}{\left(1 + \sqrt{1+\delta} \right)^{1/3}}
    \label{eq:Minimum radius - mu-scheme}
\end{eqnarray}
where $\delta=\Delta/(27 M^{2})$, such that
$x_{\Delta}(M)\approx(2M\Delta)^{1/3}$ for $M\gg\sqrt{\Delta}$. This
expression can be thought of as the minimal areal radius for the vacuum
solution which marks a reflection-symmetry surface \footnote{From
    now on, we refer to the transition surface by this name to be consistent
    with the conventions of \cite{Idrus1}.} joining the black hole to a white
hole; see \cite{Idrus1} for details. (Qualitatively, these properties are
similar to the case of constant $\lambda$ given earlier in
\cite{alonso2022nonsingular,BBVeffBH}; see also \cite{MassCovariance}.)

A massless test scalar field can be minimally coupled to the holonomy-modified
system such that its propagation follows the standard Klein--Gordon equation
\cite{SphSymmMinCoup,Bojowald-Duque-scalar,EMGscalarQNM}
\begin{equation}\label{Min coupling KG EMG}
     \tilde{g}^{\mu\nu} \tilde{\nabla}_\mu \tilde{\nabla}_\nu \phi
     =0\,,
\end{equation}
where the derivative operator $\tilde{\nabla}$ is compatible with the modified
background metric $\tilde{g}_{\mu\nu}$ defined by (\ref{static
  Schwarzschild}). However, there are other couplings allowed by covariance,
which may be relevant for effective descriptions of new quantum-gravity
effects. We will show an example at the end of this letter. The importance of
imposing strict covariance conditions in deriving the line-element
\eqref{static Schwarzschild} in LQG, and while coupling scalar matter to it,
will also be seen in what follows. These are key new ingredients of our
construction.

\noindent {\it \textbf{Thermal distribution}:} Our first derivation determines
tunneling rates for Hawking particles in the Parikh-Wilczek picture
\cite{Parikh-Wilczek}. The metric is important because it may modify the
potential and corresponding tunneling rates. We present details for a
derivation with our emergent line-element in the appendix, but note one
crucial feature here: In the metric coefficients, $\lambda$ enters only
through the function $h(x)$, where it is multiplied by $f(x)$. The latter
function vanishes at the horizon, right where the tunneling process is
relevant. As a consequence, the standard tunneling rate is unmodified in this
specific version of an effective line-element. (Mathematically, strict
equality of the tunneling rates on classical and quantum backgrounds is
implied by the application of residues in an action integration with poles at
the horizon. The residue does not depend on $\lambda$ because this function
disappears when the expressions are evaluated at the horizon, where $f(x)=0$.)
With an unmodified tunneling rate, the standard expressions for Hawking
temperature and Bekenstein--Hawking entropy hold true as well, even in the
presence of holonomy corrections. This outcome is in contrast to
incomplete proposals of LQG effects in black-hole geometries which had not
imposed all the necessary consistency conditions \cite{Modesto, Modesto2, Borges,
  Borges2, Asier-BH-Formation}.  It corrects a long-standing misconception
\cite{Rovelli:1996dv} by showing that Hawking entropy, as measured by an
asymptotic observer, remains the same in LQG.  We now turn to the question of
which aspects of Hawking evaporation are affected by LQG effects.

Hawking \cite{Hawking} was the first to analyze the behavior of a scalar field
in the vicinity of a black hole while neglecting backreaction. The distant
past, before the cloud of dust had collapsed, and the distant future, after
the black hole has formed as seen by an asymptotic observer, are both
approximated as flat space. Matching incoming with outgoing modes through the
black-hole geometry provides the spectrum of Hawking radiation. Since the
curved geometry is position-dependent, this process happens in an effective
potential which imposes a barrier that is easier to overcome for high-frequency modes. The Hawking spectrum of the frequency-dependent number of particles
\begin{equation}\label{Hawkingdistribution2}
    \braket{N_{\omega}}=\frac{{\cal T}_{l}(\omega)}{{\rm exp}\left(\omega
        T_{\rm H}\right)-1}\,,
\end{equation}
therefore, contains not only the temperature-dependent black-body factor, but
also a greybody factor ${\cal T}_{l}(\omega)$. Unlike the former, the latter
contains information about the emergent geometry because it is sensitive to a
region around the horizon.

A test scalar field propagating in a spherically symmetric spacetime
background described by (\ref{static Schwarzschild}) can be decomposed into
radial and angular parts as
$\phi(x^{\mu})=\sum_{lm}e^{i\omega t}\Phi_{lm}(x)Y_{lm}(\theta,\varphi)$, with
spherical harmonics \( Y_{lm}(\theta, \phi) \).  Equation~(\ref{Min
  coupling KG EMG}) then implies that the radial modes $\Phi(x)$ satisfy the
equation
\begin{equation}\label{Master 2}
    \left[\frac{{\rm d}^2}{{\rm d}x_{*}^{2}}+U_{l}(x)\right]\left(x\Phi_{lm}\right)=0
  \end{equation}
where $x_{*}$ is the radial tortoise coordinate defined by ${\rm d}x_{*}=
f(x)^{-1} h(x)^{-1/2} {\rm d}x$, and 
\begin{eqnarray}\label{eq:Scalar effective potential}
    &&\!\!U_{l}(x)=\omega^{2}-V_{l}(x) \nonumber \\
    &&=\omega^{2}-\left(1-\frac{2M}{x}\right)\left[\frac{2M}{x^3}+\frac{l(l+1)}{x^2}\right.\\
    &&\left.+\left(1-\frac{2M}{x}\right)\left(\frac{3M\lambda(x)^2}{x^3}+\frac{\lambda(x)\lambda'(x)}{x}\left(1-\frac{2M}{x}\right)\right)\right]\nonumber 
\end{eqnarray}
is the effective potential.
Equation~(\ref{Master 2}) has a complicated solution, but it can still provide
analytical information about physical effects in the asymptotic region. As
before, compared with the classical expression, all $\lambda$-terms are multiplied
by additional factors of $1-2M/x$ that vanish at the horizon. Nevertheless,
greybody factors depend on $\lambda$ because they are sensitive to the entire potential.

Analytical computations of greybody factors are not feasible \cite{Page,
  Harlow} which complicates the precise evaluation of the evaporation
process. We resort to the standard approximation of low-frequency modes in which
we assume that the incoming scalar particle's energy is below the background
thermal energy, $ \omega\ll T_{\rm H}$ or $M\omega \ll 1$, and expand
the distribution in terms of the latter quantity. This step allows us to extract
physical consequences.

In order to apply this approximation \cite{Unruh2, Neitzke}, we need to
identify three regions: \textit{i)} close to the horizon where the potential is
nearly vanishing, \(V_l(x) \ll \omega^2\), \textit{ii)} the intermediate
region near the peak of the potential where \(V(x) \gg \omega^2\), and
\textit{iii)} the asymptotic region where the potential flattens out again.
By solving Eq.~(\ref{Master 2}) in these three regions and requiring
continuity at the boundaries, one obtains the wave function parametrized by
the coefficient of the first region. The part of the wavefunction that
traverses the potential and enters the black hole can be quantitatively computed by
evaluating the transmission coefficient
\begin{equation}
    {\cal T}_{0}(\omega)=\frac{J_{\rm H}}{J_{\rm in}}=\frac{16M^{2}\omega^{2}}{\left(1-4M^{2}\omega^{2}\right)^{2}+4M^{2}\omega^{2}\left(1+
    \frac{\Delta}{12M^{2}}\right)^{2}}\nonumber
\end{equation}
where \(
J_{\rm in} \) and \( J_{\rm H} \) denote the complete $s$-wave flux 
currents from the incident and transmitted waves, respectively.  We recover universality of the black-hole absorption rate
\cite{Mathur, Harmark}, which to leading order in $M\omega$ depends solely on the horizon area (without the need to impose the classical limit $\Delta \rightarrow 0$).

\begin{figure*}[ht!]
    \begin{subfigure}{0.35\textwidth}
        \centering
        \includegraphics[width=1\textwidth]{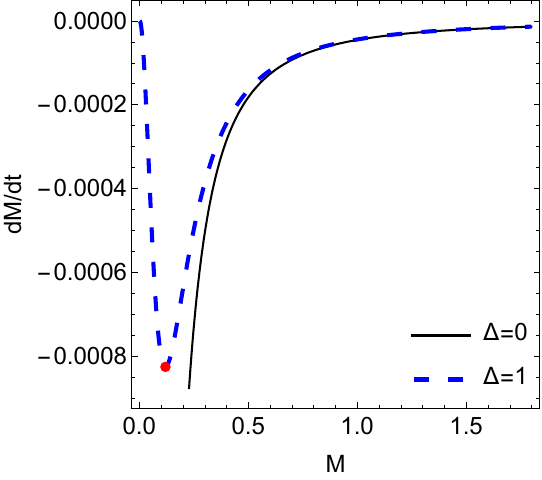}
        \caption{}
        \label{Fig: Emission rates}
    \end{subfigure}
    \hfill
    \begin{subfigure}{0.3\textwidth}
        \centering
        \includegraphics[trim=6cm 11.6cm 6cm 11.9cm,clip=true,width=0.85\textwidth]{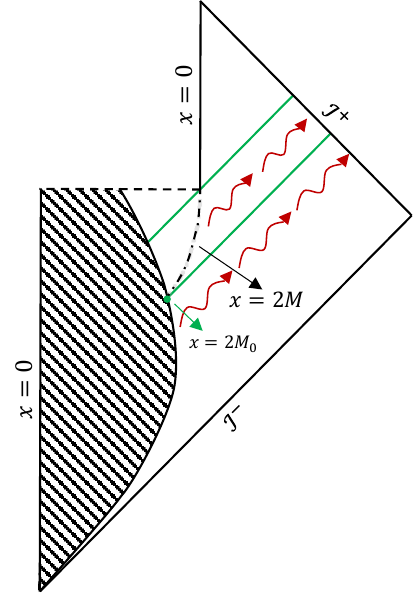}
        \caption{}
        \label{fig:image1}
    \end{subfigure}
    \hfill
    \begin{subfigure}{0.3\textwidth}
        \centering
        \includegraphics[trim=6cm 10.3cm 5cm 11cm,clip=true,width=0.83\textwidth]{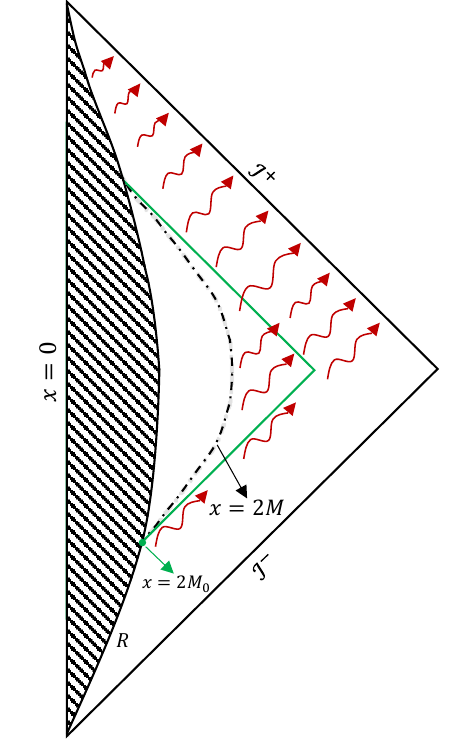}
        \caption{}
        \label{fig:BtWH}
    \end{subfigure}
    \caption{(a) Energy emission for a classical black hole (solid line) and the holonomy corrected one (dashed line) near Planckian scales.
    (b) Classical collapse and evaporation: the black hole features a singularity denoted by the dashed line and evaporation ends in an empty Minkowski space. (c) Black-to-white-hole transition. The apparent horizon $x=2M(t)$ is timelike due to energy loss; in the early stages it remains near null, but as energy emissions grow it deviates prominently into a timelike trajectory from which matter can escape if not met with a singularity.}
    \label{fig:collapsing}
\end{figure*}

\noindent {\it \textbf{Emission rate}:}
Holonomy corrections modify the transmission amplitudes and, consequently,
impact the energy emission rate
\begin{eqnarray}\label{eq:Energy emission}
    \frac{{\rm d}M}{{\rm d}t}&=&-\frac{1}{2\pi}\int_{0}^{\infty}{\rm d}\omega \frac{\omega \mathcal{T}_{0}(\omega)}{{\rm exp}(8\pi M\omega)-1}\,. 
\end{eqnarray}
If general relativity was valid throughout the entire process, the black
hole’s energy would eventually be fully depleted, returning to flat Minkowski
spacetime (Fig.~\ref{fig:image1}). Holonomy corrections from non-zero
$\Delta$ enter as a sub-leading term that slows down the emission rate as the hole evaporates,
suggesting that quantum-gravity corrections alter the final stage of a black
hole compared with the classical theory. We arrive at this conclusion without
setting a Planckian cut-off by hand, either in the momentum integral
\cite{Ivan1, Jacobson} (which would
alter the emission rate by violating Lorentz invariance) or in the two-point function \cite{Ivan2}. Instead, our result
follows naturally from a covariant description of non-perturbative effects in
LQG.

As shown in Fig.~\ref{Fig: Emission rates}, quantum spacetime effects alter
the energy emission rates of the black hole. While a classical black hole
evaporates perpetually with an accelerating rate of energy emission, the new
holonomy effect eventually slows down the evaporation process, around the
sub-Planckian mass $M_r\approx0.12\sqrt{\Delta}$. After reaching this mass, the
emission rate decreases sharply and its Hawking radiation slows down
remarkably as compared to that of a classical black-hole, achieving a
vanishing mass only in an infinite amount of time. Quantum effects can
therefore stabilize a black hole thermally which, taken at face-value, seems
to indicate the formation of a remnant.

Qualitatively, this outcome is not unexpected. Vacuum spherically-symmetric
solutions of effective LQG equations have been used to show that, near the
Planck scale, the classical singularity is replaced by a reflection-symmetry
surface. The eternal solution (\ref{static Schwarzschild}) suggests that once
matter passes through this surface, it `bounces' into a white hole structure
\cite{alonso2022nonsingular, BBVeffBH, Idrus1}. However, this vacuum
description does not determine how the process of collapsing matter is
perceived by a distant observer. Matter could bounce into a white hole which
shares the same asymptotic region as the black hole and thus be visible to an
observer, or split off as a remnant with a disconnected asymptotic region. The
latter case would be described as the formation of an interuniversal wormhole
\cite{EMGPF,Idrus1}, and the former as a black-to-white-hole transition
\cite{Carlo1,Haggard}. As we show in what follows, backreaction of Hawking
radiation plays a central role in resolving this ambiguity of outcomes. We
discuss the two potential final stages of the black hole, one described by a
thermally stabilized remnant, and the other by a transition
process from a black hole to a white hole; see Fig.~\ref{fig:collapsing}.

\noindent {\it \textbf{The evaporation process}:} 
The emission rate slows down at a sub-Planckian mass $M_r$, suggesting a stable
remnant. However, once in this quantum space-time regime, the
horizon area can only be resolved up to quantum
  fluctuations. Hawking radiation then ends when the horizon
  area reaches, within quantum fluctuations, the smallest possible radius
  $x_{\Delta}(M)$. Based on the uncertainty relation, quantum fluctuations
should be proportional to $\sqrt{\hbar}\propto \ell_{\rm Pl}$, with an
additional factor of $M$ required for the correct dimensions of area
fluctuations. Following \cite{Parikh:2024zmu}, a specific expression for area
fluctuations is given by $2 M \sqrt{8 \pi \Delta}$, where we have replaced
$\ell_{\rm Pl}$ by $\sqrt{\Delta}$ as the fundamental UV-scale in
LQG. Upon equating the fluctuating horizon radius with the size
  of the minimum-area surface, $x_{\Delta}(M)$, we find
  $M_f \approx 3.8 \sqrt{\Delta}$, a value much bigger than the mass $M_r$
  when the emission rate starts decreasing to initiate the formation of the
  remnant. The dissolving horizon implies an end of Hawking evaporation at $M_f$, rather than $M_r$.

  Backreaction can be taken into account by adjusting the
  black-hole geometry to decreasing masses according to (\ref{eq:Energy
    emission}). When the evaporation process ends at $M_f$, the
  continuum spacetime geometry, and therefore the classical picture of a
  horizon-bound black hole, dissolves. Instead, we have a superposition of
  a horizonless object and the final stage of the collapsing
  black-hole. Semiclassically, this quantum state and its future gives a
  coarse-grained description of a transition from a black hole to
  its opened-up interior. Importantly, this stage is approached in finite time, before a
  remnant could form.

  In this process, energy of the collapsing matter leaks out after some time
  as described in Fig.~\ref{fig:BtWH}. A final observation made in
  \cite{EMGscalarQNM} allows us to distinguish between a wormhole outcome and
  a white hole: Some quasi-normal modes of a non-minimally coupled scalar
  field--- derived from the same background (\ref{static Schwarzschild})---
  can become unstable for Planckian black holes at the critical mass
  $M_c\approx0.57\sqrt{\Delta}>M_r$, jump-starting the white-hole emission. We
  obtain the same thermal stabilization for Planckian black holes for a
  non-minimally coupled scalar field, as shown in the appendix, but it
  destabilizes gravitationally. This outcome can be understood as an
  implication of a repulsive quantum-gravity contribution, which helps to
  avoid the singularity and can be relevant for Planckian black holes in their
  exterior too. A quantum black hole is no longer gravitationally bound and
  therefore transitions to a white hole. This instability does not happen for
  minimal coupling, but insisting on such a choice within a general effective
  theory of quantum gravity would amount to fine-tuning. Even if non-minimal
  coupling coefficients are small, the instability of some modes ensures that
  their implications are relevant. Instead of thermal Hawking radiation, the
  final stages of black-hole evaporation are dominated by a gravitational
  instability, subject to unitary quantum evolution.

  \noindent {\it \textbf{Conclusions}:} As the mass of an evaporating LQG
  black hole shrinks toward $M_r$, holonomy corrections dramatically slow down
  Hawking radiation. When the horizon area reaches a Planckian size, quantum
  fluctuations can no longer be ignored. Within fluctuations, the horizon
  reaches the minimum-area reflection-symmetry surface at $M_f>M_r$. In the
  underlying quantum state, there is a finite probability for a sub-Planckian
  black hole to tunnel into a white hole \cite{Carlo-Bianchi}, an outcome that
  would always be preferred given that it takes a long time to form the
  remnant at mass $M_r$. Going beyond heuristic claims, we provide a physical
  mechanism, based on a holonomy-induced instability in a quasi-normal mode
  which constitutes an explicit candidate for an initial white-hole emission at a
  mass scale $M_c>M_r$. Crucially, the mass hierarchy $M_r<M_c<M_f$ derived
  here reveals that the black hole first turns quantum, as shown by
  significant quantum fluctuations at $M_f$, and forms a superposition of a
  horizon-covered black hole with an exposed white hole.  This stage is
  followed by the quantum-gravitational instability triggering an explosion at
  $M_c$, well before Hawking radiation starts weakening at $M_r$.

  Our results provide the first covariant, self-consistent description of
  black-hole evaporation in LQG with several crucial ingredients of quantum
  spacetime falling into the right places. Compared to previous attempts, we
  incorporated two new contributions with important implications for the
  physical spacetime of LQG black holes: consistent matter couplings and
  backreaction effects. Computing Hawking radiation for non-dynamical vacuum
  models initially suggests the formation of a stable remnant. However, upon
  factoring in backreaction and fluctuation effects, we find that the black
  hole transitions to a white hole before the remnant mass is reached, enabled
  by new (quantum) gravitational instabilities at high density.  Thus,
  dynamical evaporation is essential to not only differentiate between
  competing scenarios, but also in establishing the final stage of black holes
  in quantum gravity. The inclusion of backreaction is possible thanks to the
  singularity-free nature of our dynamical solutions. In this picture,
  information inside the black hole naturally flows into the white hole, but
  only once the mass of the evaporating black hole is comparable to Planckian
  scales where LQG effects become large and backreaction cannot be ignored.

\section{Acknowledgments}
IHB is supported by the Indonesia Endowment Fund for Education (LPDP) grant from the Ministry of Indonesia. The work of MB and EID was supported in part by NSF grant PHY-2206591. SB is supported in part by the Higgs Fellowship and by the STFC Consolidated Grant “Particle Physics at the Higgs Centre”. SB acknowledges support from the Blaumann Foundation for a grant on ‘Black Holes in Loop Quantum Gravity’.

\appendix
\section*{Appendix}
 
{\bf \textit{Hawking Radiation}:} Visualizing Hawking radiation as
energy-conserving particle tunneling  \cite{Parikh-Wilczek}, the energy
at infinity (the ADM mass) can be written as  $\tilde{H}=\omega+\left(M_{\rm
    ADM}-\omega\right)$. The first term is the kinetic energy of the
outgoing particle while the second term is the change in the black hole's
potential energy. For our choice of a decreasing holonomy function, the ADM
mass equals the black hole's mass parameter $M_{\rm ADM}=M$
\cite{Idrus1}. (Note that this is no longer true for generic $\lambda(x)$,
where $M_{\rm ADM}$ may differ from $M$ \cite{Idrus3}.) Subsequently, the
particle follows a geodesic determined by the line-element (\ref{static Schwarzschild}), or
\begin{eqnarray}\label{PG metric}
    &&{\rm d}s^2=-{\rm d}t_{\rm PG}^2+x^2{\rm d}\Omega^2
    \\
    &&\quad
    + \frac{\left({\rm d}x+\sqrt{\frac{2(M-\omega)}{x}}\sqrt{1+\lambda(x)^2\left(1-\frac{2(M-\omega)}{x}\right)}\;{\rm d}t_{\rm PG}\right)^2}{\left(1+\lambda(x)^2\left(1-\frac{2(M-\omega)}{x}\right)\right)},
    \nonumber
\end{eqnarray}
expressed in Painlev\'e--Gullstrand coordinates. The tunneling rate is then
computed using the Wentzel--Kramers--Brillouin (WKB) approximation. It takes the
form \(\Gamma \simeq \exp\left(-2\mathrm{Im} [S]\right)\) where \(S\)
represents the on-shell action for an \(s\)-wave-particle traveling from
\(x_{\rm H}^{\mathrm{in}} = 2\left(M - \omega\right) - \epsilon\) to \(x_{\rm
  H}^{\mathrm{out}} = 2\left(M - \omega\right) + \epsilon\). Using a point-particle energy effect on the metric is justified due to the significant blue-shift near the horizon.

Adapting \cite{Parikh-Wilczek}, the on-shell action for an $s$-wave traveling from just within the horizon to just outside the horizon can be expressed as
\begin{equation}
S=\int^{\omega}_{0}{\rm d}\omega'
\int_{x_{\rm H}^{\mathrm{in}}}^{x_{\rm H}^{\mathrm{out}}} \frac{\left(1-\sqrt{\frac{2\left(M-\omega'\right)}{x}}\right)^{-1} {\rm d}x}{\left(1+\lambda(x)^{2}\left(1-\frac{2\left(M-\omega'\right)}{x}\right)\right)}\,.
\end{equation}
This expression is integrated using the $i\epsilon$-prescription with the poles located
at the horizon. Consequently, the residue does not receive holonomy
corrections, yielding the tunneling rate
\begin{equation}\label{eq:tunneling rate}
    \Gamma\simeq{\rm exp}\left[-8\pi\omega \left(M-\frac{\omega}{2}\right)\right]\,.
\end{equation}
Therefore, we recover the Hawking distribution (\ref{Hawking distribution})
with an effective temperature
\begin{equation}
  T_{\rm H}=1/\left(8\pi \left(M-\frac{\omega}{2}\right)\right)\,.
\end{equation}
The way $\lambda(x)$ shows up in the line-element \eqref{static
  Schwarzschild}, which follows from covariance conditions \cite{Idrus1}, is
crucial in ensuring that the classical result continues to hold in LQG. The
argument inside the exponent of (\ref{eq:tunneling rate}) yields the change in
the black hole entropy as measured by an asymptotic observer,
\begin{equation}
    S_{\infty}=    S_{\rm BH}\,,
\end{equation}
where $S_{\rm BH}=A_{\rm H}/4$ is the Bekenstein-Hawking entropy, and
$A_{\rm H}=8\pi M^2$ the horizon's area. This result is consistent with
deriving entropy from the quasilocal Brown--York energy for this choice of
holonomy function \cite{Idrus1}.

The spectrum of Hawking radiation is derived by matching incoming and outgoing
modes in asymptotic regions. Once the black hole reaches its
equilibrium state, the initial Minkowski vacuum modes,
\( u_{\rm in}(\omega) \), defined at past null infinity (\( \mathscr{I}^- \)),
are perceived as a thermal state by an observer at future null infinity
(\( \mathscr{I}^+ \)).  The completeness of the modes at \( \mathscr{I}^- \),
implies that the outgoing modes, \( u_{\rm out}(\omega) \), that reach
\( \mathscr{I}^+ \) can be expanded in terms of the original basis. The
geometric optics approximation then establishes the Bogulibov transformation
connecting \( \mathscr{I}^- \) to \( \mathscr{I}^+ \).

We consider a collapsing system that forms a stationary black hole with a test
scalar field in a Minkowski vacuum state $\ket{0_{\rm in}}$ at past infinity.
The exterior region is described by the line-element (\ref{static
  Schwarzschild}), which exhibits timelike Killing symmetry that ensures a
unique vacuum state $\ket{0_{\rm out}}$ at future null infinity. Quantization
can be performed in both asymptotic regions starting in a vacuum state
$\ket{0_{\rm in}}$, defined by the complete basis
$\{f_{\omega},f^{*}_{\omega}\}$ at $\mathscr{I}^{-}$, and ending in another
vacuum state $\ket{0_{\rm out}}$, defined by the basis
$\{g_{\Omega},g^{*}_{\Omega}\}$ at $\mathscr{I}^{+}$.

The early modes $f_{\omega}$ are prepared from $\mathscr{I}^{-}$ as 
\begin{eqnarray}
    f_{\omega}=\frac{1}{\sqrt{2\pi \omega}}e^{-i\omega v}.
\end{eqnarray}
On the other hand, the late time modes $g_{\Omega}$ propagated backward in
time would be partially reflected to give a plane wave coming from
$\mathscr{I}^{-}$ and another part would be transmitted into the horizon.
This ``tracing back in time'' technique gives the $g$-modes in the advanced
null coordinates $v$,
\begin{equation}\label{gv}
    g_{\Omega}(v)\approx \frac{1}{\sqrt{2\pi \Omega}}e^{i2\kappa^{-1}\ln(-CD v)} \,, \quad {\rm for}\quad v<0\,,
\end{equation}
where $C$ and $D$ are constant coefficients, and $\kappa=1/(4M)$ is the surface
gravity.  We express the $g$-modes in terms of the $f$-modes through the
Bogoliubov transformation
$ g_{\Omega}(v)=\int_{0}^{\infty} {\rm d}\omega
\left(A_{\Omega\omega}f_\omega+B_{\Omega\omega}f^{*}_\omega\right)$.  Applying
a Fourier transform, the Bogoliubov coefficients can be easily obtained,
\begin{equation}
    A_{\Omega\omega}=\sqrt{\frac{\omega}{\pi}}\tilde{g}_{\Omega}(\omega) \quad,\quad
    B_{\Omega\omega}=\sqrt{\frac{\omega}{\pi}}\tilde{g}_{\Omega}(-\omega)\,.
  \end{equation}
  The angular component of the test scalar field introduces a peak in the
  effective potential (\ref{eq:Scalar effective potential}), leading to a
  partial transmission coefficient ${\cal T}_{l}$ for the \textit{in}-modes,
  and a corresponding reflection coefficient ${\cal R}_{l}$.  Accordingly, the
  number of particles detected by the external observer is given by
\begin{equation}\label{Hawking distribution}
    \braket{N_{\omega}}=\frac{{\cal T}_{l}(\omega)}{{\rm exp}\left(2\pi\omega\kappa^{-1}\right)-1}\,.
\end{equation}
The transmission coefficient ${\cal T}_{l}(\omega)$ is commonly known as the greybody factor. 

The distribution (\ref{Hawking distribution}) represents the standard Hawking distribution with classical temperature $T_{\rm H}=\kappa/2\pi$.
Therefore, LQG effects modifying this distribution can only enter through the greybody factor. \\

{\bf \textit{Non-minimal coupling}:} The tortoise radial coordinate differential reads
\begin{eqnarray}
    {\rm d}x_{*}=\frac{{\rm d}x}{\left(1-\frac{2M}{x}\right)\sqrt{1+\frac{\Delta}{x^{2}}\left(1-\frac{2M}{x}\right)}}
\end{eqnarray}
and can be integrated near the horizon, which gives
\begin{eqnarray}
    x_{*}\approx x+2M\ln\left(\frac{x}{2M}-1\right)+\frac{\Delta}{2x}+\frac{3M-2x}{16x^{4}}\Delta^{2}\,.
\end{eqnarray}
In this coordinate, the nonminimal equation of motion can be reduced to a
Schr\"odinger-type equation with eigenvalue $\omega^2$,
\begin{equation}\label{eq:QNM equation - Schrodinger - WKB}
    \frac{\partial^2 S_{lm}}{(\partial x^*)^2} + \left[\omega^2-U_l(x^*)\right] S_{lm} = 0
    \,,
\end{equation}
where $U_l(x^*)$ is the effective potential
\begin{eqnarray}\label{eq:Potential - Schrodinger}
    U_l = V_l + \zeta^2- \frac{\partial \zeta}{\partial x^*}
\end{eqnarray}
with
\begin{eqnarray}
    &&V_l (x) = \frac{h(x)}{x^2} \left(1-\frac{2 M}{x}\right) \left(l(l+1)+\frac{2M}{x}\right)
    \,,\\
    &&\zeta = - \frac{\lambda^2}{2 \sqrt{h(x)}} \left(1-\frac{2 M}{x}\right) \left( \frac{2 M}{x^2}
    + \left(1-\frac{2 M}{x}\right) \frac{\partial\ln \lambda^2}{\partial x} \right)
    \,,\nonumber\\\label{eq:Damping term - general}
\end{eqnarray}
and $S_{lm}$ is the modulated amplitude defined by
\begin{equation}\label{eq:Fourier-modulation relation}
    \tilde{u}_{lm} (\omega,x^*) = Z(x^*) S_{lm} (\omega,x^*)
\end{equation}
with
\begin{equation} \label{modulation}
    Z (x^*) = \exp \left( - \int{\rm d} x^*\; \zeta \right)=\sqrt{1-\frac{2M \Delta}{x^{3}}+\frac{\Delta}{x^{2}}}
    \,.
\end{equation}

The flux current is given by
\begin{eqnarray}
     j^{x}=\frac{i}{2}\frac{N\sqrt{f(x)}}{Z^{2}}\left[f(x)\left(u_{lm}\partial_xu_{lm}^{*}-u_{lm}^{*}\partial_xu_{lm}\right)\right]\,.
\end{eqnarray}
Since $Z(\infty)=Z(2M)=1$, the total number of particles that are reflected plus those transmitted is conserved (unlike in the case of a constant holonomy function \cite{Idrus3}).

The transmission coefficient reads
\begin{eqnarray}
     \mathcal{T}_{0}(\omega)\simeq \frac{16M^{2}
\omega^{2}}{\left(1-4M^{2}\omega^{2}\right)^{2}+4M^{2}\omega^{2}\left(1+\frac{\Delta}{8M^{2}}\right)^{2}}
\end{eqnarray}
\begin{figure}[h]
    \centering
\includegraphics[width=7cm]{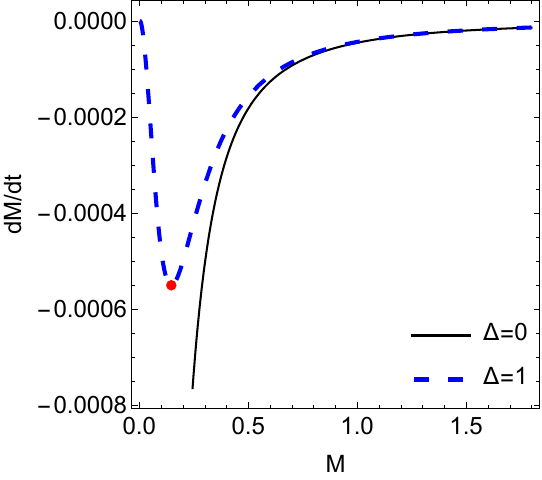}
    \caption{Energy emission rate for $\Delta=1$ and non-minimal coupling of
      the scalar field. The red dot corresponds to a remnant mass of $M_r \approx 0.15$.}
    \label{Fig: NMC}
\end{figure}
Using the relation (\ref{eq:Energy emission}), we can then plot the amount of energy flux in the non-minimally coupled case. The plot of the evaporation is shown in Fig. (\ref{Fig: NMC}). The remnant mass appears at $M_r\approx 0.15\sqrt{\Delta}$. \\

{\bf\textit{Quasinormal modes}:} 
Setting the standard boundary conditions at the horizon and asymptotically,
\begin{eqnarray}
    \label{eq:Boundary cond - horizon}
    S_{lm} (\omega,x^*) &\approx& e^{- i\omega x^*}
    \approx \left(x-2M\right)^{- i x_{\rm H}\omega}
    \,\,,\, x^* \to - \infty
    \,,\nonumber \\
    \label{eq:Boundary cond - Inf}
    S_{lm} (\omega,x^*) &\approx& e^{+ i\omega x^*}
    \approx x^{i\omega x_{\rm H}} e^{+ i\omega x}
   \,\,,\, x^* \to + \infty
\end{eqnarray}
the equation of motion (\ref{eq:QNM equation - Schrodinger - WKB}) can be  treated with the third-order WKB approximation and the frequency modes can be obtained numerically \cite{EMGscalarQNM}.

\begin{figure}[h]
        \centering
        \includegraphics[width=0.4\textwidth]{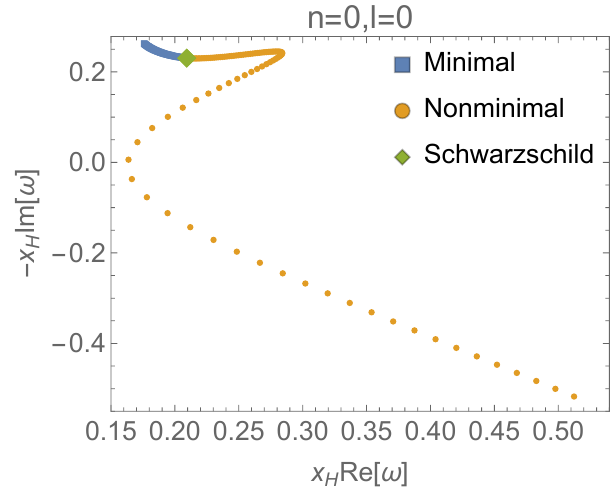}
        \caption{Positive branch of the eigenfrequency mode $\omega_{00}$. The parameter $\Delta/x_{\rm H}^2$ is varied from $0$ to $1$. The imaginary part of the eigenfrequency vanishes at $\Delta/x_{\rm H}^2\approx 0.7714$.}
        \label{fig:w00mubar}
\end{figure}

In particular, the imaginary part of the spectrum of the $\omega_{00}$ mode, shown in Fig.~\ref{fig:w00mubar}, continuously goes through zero changing sign at the parameter value $\Delta/x_{\rm H}^2\approx 0.77142$, corresponding to the critical mass $M_c\approx 0.57\sqrt{\Delta}$.
In contrast, its real part remains finite and positive.
This behavior indicates that this mode is potentially destabilizing: While a positive imaginary part of the frequency leads to a decaying mode, a negative one leads to an amplifying mode.
Since the energy to amplify the waves can only come from the black hole, this provides the mechanism for a final explosion into a white hole.

\pagebreak
\end{document}